\documentclass{article}
\usepackage{spconf,amsmath,graphicx}

\usepackage{times}
\usepackage{epsfig}
\usepackage{amssymb}
\usepackage{microtype}
\usepackage{booktabs}
\usepackage[caption=false,font=footnotesize]{subfig}
\usepackage{pifont}
\usepackage{relsize}
\usepackage{multirow}
\usepackage{gensymb}
\usepackage{tabularx}
\usepackage{comment}
\usepackage{xcolor}
\usepackage{amsmath}
\usepackage{cite}
\usepackage{etoolbox}

\newcommand{\cmark}{\text{\ding{51}}}
\newcommand{\xmark}{\text{\ding{55}}}

\title{End-to-End Multi-channel Transformer for Speech Recognition}

\name{Feng-Ju Chang, Martin Radfar, Athanasios Mouchtaris, Brian King, and Siegfried Kunzmann}
\address{Alexa Machine Learning, Amazon, USA \\
\tt\{fengjc, radfarmr, mouchta, bbking, kunzman\}@amazon.com}

\begin{document}
\ninept
\maketitle
\begin{abstract}
Transformers are powerful neural architectures that allow integrating different modalities using attention mechanisms. In this paper, we leverage the neural transformer architectures for multi-channel speech recognition systems, where the spectral and spatial information collected from different microphones are integrated using attention layers. Our multi-channel transformer network mainly consists of three parts: channel-wise self attention layers (CSA), cross-channel attention layers (CCA), and multi-channel encoder-decoder attention layers (EDA). The CSA and CCA layers encode the contextual relationship ``within" and ``between" channels and across time, respectively. The channel-attended outputs from CSA and CCA are then fed into the EDA layers to help decode the next token given the preceding ones.
The experiments show that in a far-field in-house dataset, our method outperforms the baseline single-channel transformer, as well as the super-directive and neural beamformers cascaded with the transformers.
\end{abstract}

\section{Introduction}
\label{sec:intro}

In the past few years, voice assisted devices have become ubiquitous, and enabling them to recognize speech well in noisy environments is essential. One approach to make these devices robust against noise is to equip them with multiple microphones so that the spectral and spatial diversity of the target and interference signals can be leveraged using beamforming approaches~\cite{omologo2001speech,wolfel2009distant,kumatani2012microphone,kinoshita2016summary,virtanen2012techniques,menne2016rwth}. It has been demonstrated in \cite{barker2015third,kinoshita2016summary,menne2016rwth} that beamforming methods for multi-channel speech enhancement produce substantial improvements for ASR systems; therefore, existing ASR pipelines are mainly built on beamforming as a pre-processor and then cascaded with an acoustic-to-text model~\cite{wolfel2009distant,chang2019mimo,chang2020end,kumatani2019multi}.

A popular beamforming method in the field of ASR is super-directive (SD) beamforming~\cite{doclo2007superdirective,himawan2010clustered}, which uses the spherically isotropic noise field and computes the beamforming weights. This method requires the knowledge of distances between sensors and white noise gain control~\cite{wolfel2009distant}.
With the great success of deep neural networks in ASR, there has been significant interest to have end-to-end all-neural models in voice assisted devices. Therefore, neural beamformers are becoming state-of-the-art technologies for the unification of all-neural models in speech recognition devices ~\cite{heymann2016neural,erdogan2016improved,ochiai2017multichannel,chang2019mimo,chang2020end,minhua2019frequency,kumatani2019multi,li2016neural,xiao2016deep,meng2017deep}. In general, neural beamformers can be categorized into fixed beamforming (FBF) and adaptive beamforming (ABF). 
While the beamforming weights are fixed in FBF \cite{minhua2019frequency,kumatani2019multi} during inference time, the weights in adaptive beamforming (ABF)
~\cite{heymann2016neural,erdogan2016improved,ochiai2017multichannel,chang2019mimo,chang2020end,li2016neural}, can vary based on the input utterances~\cite{li2016neural,meng2017deep} or the expected speech and noise statistics computed by a neural mask estimator~\cite{heymann2016neural,erdogan2016improved} and the well-known MVDR formalization~\cite{capon1969high}.

Transformers~\cite{vaswani2017attention} are powerful neural architectures that lately have been used in ASR~\cite{dong2018speech,lu2020exploring,wang2020transformer}, SLU~\cite{radfar2020end}, and other audio-visual applications~\cite{paraskevopoulos2020multiresolution} with great success, mainly due to their attention mechanism. Only until recently, the attention concept has also been applied to beamforming, specifically for speech and noise mask estimations~\cite{tolooshams2020channel,chang2020end}.
While theoretically founded via MVDR formalization~\cite{capon1969high}, a good speech and noise mask estimator needs to be pre-trained on synthetic data for the well-defined target speech and noise annotations; the speech and noise statistics of synthetic data, however, may be far away from real-world data, which can lead to noise leaking into the target speech statistics and vice-versa~\cite{drude2019unsupervised}. This drawback could further deteriorate its finetuning with the cascaded acoustic models.

In this paper, we bypass the above front-end formalization and propose an end-to-end multi-channel transformer network which takes directly the spectral and spatial representations (magnitude and phase of STFT coefficients) of the raw channels, and use the attention layers to learn the contextual relationship within each channel and across channels, while modeling the acoustic-to-text mapping. The experimental results show that our method outperforms the other two neural beamformers cascaded with the transformers by 9\% and 9.33\% respectively, in terms of relative WER reduction on a far-field in-house dataset. 
In Sections~\ref{sec:method}, \ref{sec:exps}, and \ref{sec:conc}, we will present the proposed model, our experimental setup and results, and the conclusions, respectively. 

\begin{figure*}[t]
\centering
\includegraphics[width=0.75\textwidth]{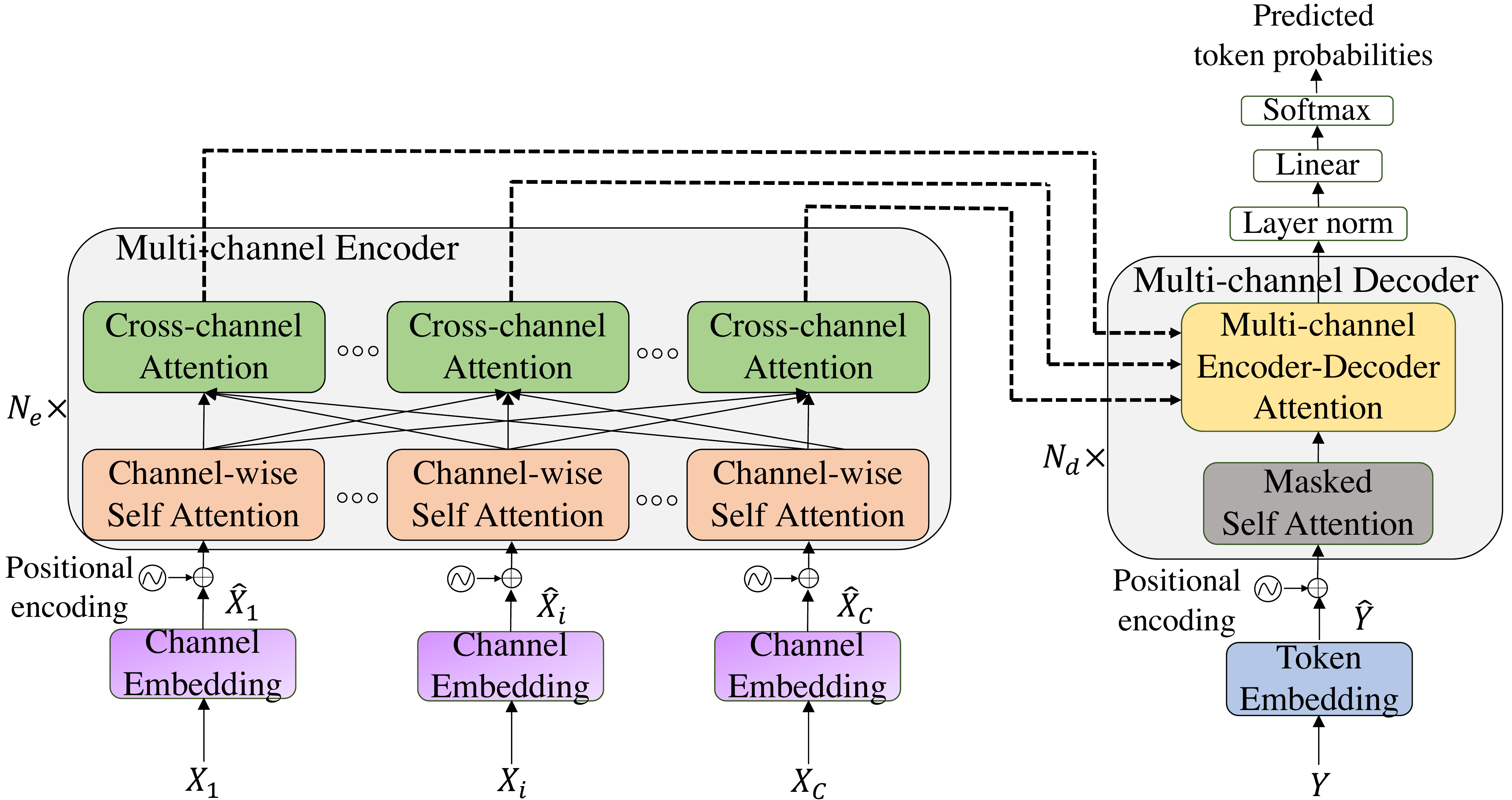}
\caption{An overview of the proposed multi-channel transformer network. $C$, $N_e$, and $N_d$ are the number of channels, encoder layers and decoder layers, respectively. Note that the audio sequences $X_1$,...,$X_i$,...,$X_C$ share the same token sequence $Y$.}
\vspace{-4mm}
\label{fig:model_diagram}
\end{figure*}

\section{Proposed Method}
\label{sec:method}

\begin{figure*}[t]
\centering
\includegraphics[width=0.75\textwidth]{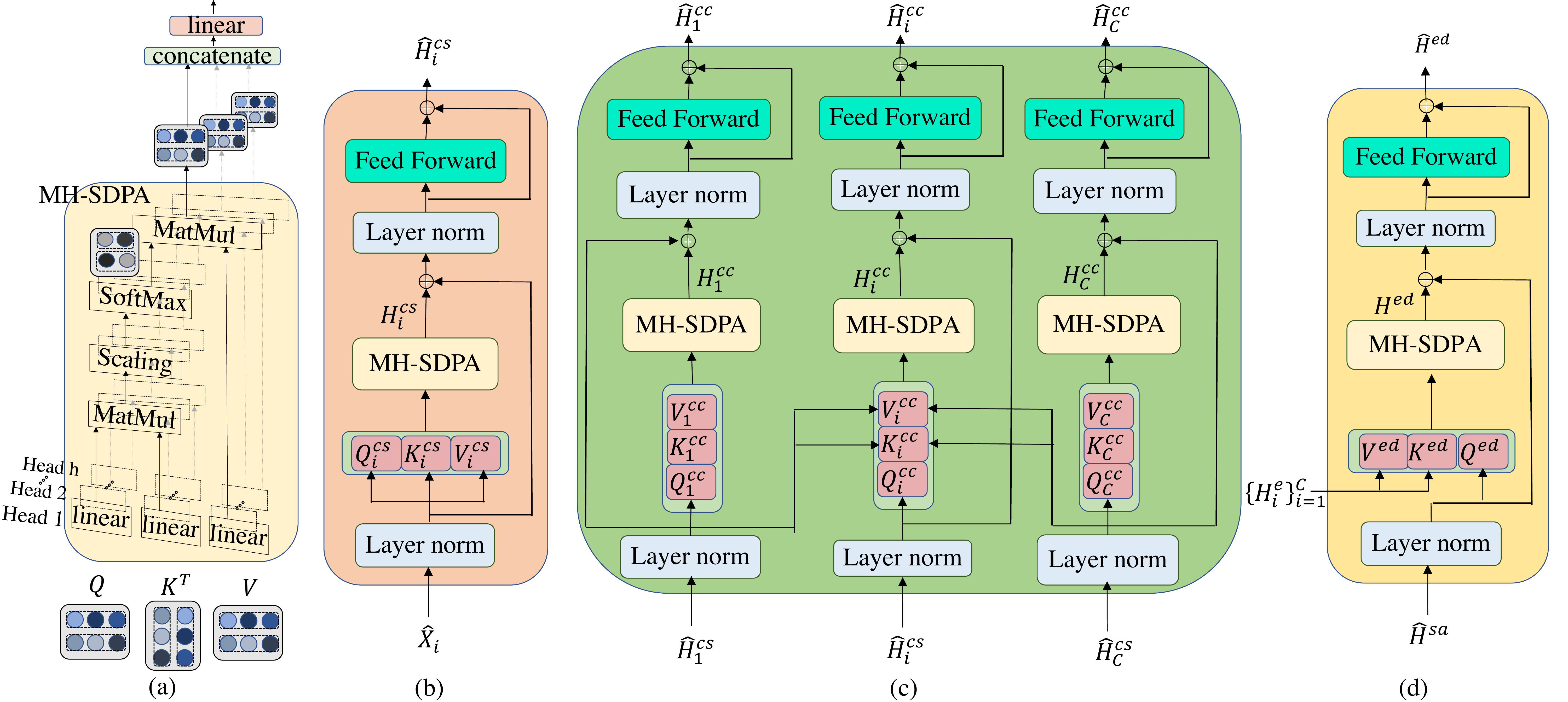}
\caption{The attention blocks in our multi-channel transformer. (a) shows the multi-head scaled dot-product attention (MH-SDPA). (b), (c), (d) show a channel-wise self attention layer (CSA), a cross-channel attention layer (CCA), and a multi-channel encoder-decoder attention layer (EDA) respectively. }
\vspace{-4mm}
\label{fig:attentions}
\end{figure*}

Given $C$-channels of audio sequences $\mathcal{X}=(X_1,...,X_i,...,X_C)$ and the target token sequence $\mathcal{Y}=(\textbf{y}_1,...,\textbf{y}_j,...,\textbf{y}_U)$ with length $U$, where $X_i \in \mathbb{R}^{T \times F}$ is the $i^{th}$-channel feature matrix of $T$ frames and $F$ features, and $\textbf{y}_j \in \mathbb{R}^{L \times 1}$ is a one-hot vector of a token from a predefined set of $L$ tokens, our objective is to learn a mapping in order to maximize the conditional probability $p(\mathcal{Y}|\mathcal{X})$. An overview of the multi-channel transformer is shown in Fig.~\ref{fig:model_diagram}, which contains the channel and token embeddings, multi-channel encoder, and multi-channel decoder. For clarity and focusing on how we integrate multiple channels with attention mechanisms, we will omit the multi-head attention~\cite{vaswani2017attention}, layer normalization~\cite{ba2016layer}, and residual connections~\cite{he2016deep} in the equations, but only illustrate them in Fig.~\ref{fig:attentions}. 

\subsection{Channel and Token Embeddings}
Like other sequence-to-sequence learning problems, we start by projecting the source channel features and one-hot token vector to the dense embedding spaces, for more discriminative representations. The $i^{th}$ channel feature matrix, $X_i$, contains magnitude features $X_i^{mag}$ and phase features $X_i^{pha}$; more details will be described in Sec.~\ref{sec:exps}. We use three linear projection layers, $W^{me}_{i}$, $W^{pe}_{i}$, and $W^{je}_{i}$ to embed the magnitude, phase, and their concatenated embeddings, respectively. Since the transformer networks do not model the position of a token within a sequence, we employ the positional encoding ($\textbf{PE}$)~\cite{vaswani2017attention} to add the temporal ordering into the embeddings. The overall embedding process can be formulated as:
\begin{align}
    \hat{X_i} = [X_i^{mag} W^{me}_{i}, X_i^{pha} W^{pe}_{i}] W^{je}_{i} + \textbf{PE}(t,f)
    \label{eq:channel_embedding}
\end{align}
Here, all the bias vectors are ignored and $[.,.]$ indicates the concatenation. $\hat{X_i} \in \mathbb{R}^{T \times d_m}$, where $d_m$ is the embedding size. $i \in \{1,...,C\}$, $t \in \{1,...,T\}$, and $f \in \{1,...,d_m\}$. Similarly, the token embedding is formulated as:
\begin{align}
    \hat{\textbf{y}}_{j} = W^{te} \textbf{y}_{j} + \textbf{b}^{te} + \textbf{PE}(j,l)
    \label{eq:token_embedding}
\end{align}
Here $W^{te}$ and $\textbf{b}^{te}$ are learnable token-specific weight and bias parameters. $\hat{\textbf{y}}_{j} \in \mathbb{R}^{d_m \times 1}$, $j \in \{1,...,U\}$, and $l \in \{1,...,d_m\}$.

\subsection{Multi-channel Encoder}
\textbf{Channel-wise Self Attention Layer (CSA)}: Each encoder layer starts from utilizing self-attention layers per channel (Fig.~\ref{fig:attentions}(b)) in order to learn the contextual relationship within a single channel. Following~\cite{vaswani2017attention}, we use the multi-head scaled dot-product attention (MH-SDPA) as the scoring function shown in Fig.~\ref{fig:attentions}(a) to compute the attention weights across time. Given the $i^{th}$ channel embeddings, $\hat{X}_i$, by Eq.(\ref{eq:channel_embedding}), we can obtain the queries, keys, and values via the linear transformations followed by an activation function as:
\begin{align}
    Q^{cs}_{i}=\sigma\left(\hat{X}_i W^{cs,q} + \textbf{1} (\textbf{b}^{cs,q}_i)^T\right) \nonumber \\
    K^{cs}_{i}=\sigma\left(\hat{X}_i W^{cs,k} + \textbf{1} (\textbf{b}^{cs,k}_i)^T\right) \label{eq:cs_qkv}\\
    V^{cs}_{i}=\sigma\left(\hat{X}_i W^{cs,v} + \textbf{1} (\textbf{b}^{cs,v}_i)^T\right) \nonumber
\end{align}
Here $\sigma(.)$ is the $ReLU$ activation function,  $W^{cs,*} \in \mathbb{R}^{d_m \times d_m}$ and $\textbf{b}^{cs,*} \in \mathbb{R}^{d_m \times 1}$ are learnable weight and bias parameters, and $\textbf{1} \in \mathbb{R}^{T \times 1}$ is an all-ones vector. The channel-wise self attention output is then computed by:
\begin{align}
    H^{cs}_{i} = Softmax\left(\frac{Q^{cs}_i (K^{cs}_i)^T}{\sqrt{d_{m}}}\right) V^{cs}_i
    \label{eq:csa}
\end{align}
where the scaling $\frac{1}{\sqrt{d_{m}}}$ is for numerical stability~\cite{vaswani2017attention}. We then add the residual connection~\cite{he2016deep} and layernorm~\cite{ba2016layer} (See Fig.~\ref{fig:attentions}(b)) before feeding the contextual time-attended representations through the feed forward layers in order to get final channel-wise attention outputs $\hat{H^{cs}_i}$, as shown on the top of Fig.~\ref{fig:attentions}(b).

\textbf{Cross-channel Attention Layer (CCA)}: The cross-channel attention layer (Fig.~\ref{fig:attentions}(c)) learns not only the cross correlation in time between time frames but also cross correlation between channels given the self-attended channel representations, $\{\hat{H}^{cs}_i\}^{C}_{i=1}$. We propose to create $Q$, $K$ and $V$ as follows:
\begin{comment}
\begin{align}
    & Q^{cc}_{i}=\sigma\left(\hat{H}^{cs}_i W^{cc,q} + \textbf{1} (\textbf{b}^{cc,q}_i)^T\right) \notag \\
    & K^{cc}_{i}=\sigma\left({(\sum_{j,j\ne i} A_j \odot \hat{H}^{cs}_j)}W^{cc,k} + \textbf{1} (\textbf{b}^{cc,k}_i)^T\right) \label{eq:cc_qkv}\\
    & V^{cc}_{i}=\sigma\left((\sum_{j,j\ne i} A_j \odot \hat{H}^{cs}_j) W^{cc,v} + \textbf{1} (\textbf{b}^{cc,v}_i)^T\right) \notag
\end{align}
\end{comment}
\begin{align}
    & Q^{cc}_{i}=\sigma\left(\hat{H}^{cs}_i W^{cc,q} + \textbf{1} (\textbf{b}^{cc,q}_i)^T\right) \notag \\
    & K^{cc}_{i}=\sigma\left(H^{CCA} W^{cc,k} + \textbf{1} (\textbf{b}^{cc,k}_i)^T\right) \label{eq:cc_qkv}\\
    & V^{cc}_{i}=\sigma\left(H^{CCA} W^{cc,v} + \textbf{1} (\textbf{b}^{cc,v}_i)^T\right) \notag
\end{align}
\begin{align}
    H^{CCA} = \sum_{j,j\ne i} A_j \odot \hat{H}^{cs}_j
    \label{eq:h_cca}
\end{align}
where $\hat{H}^{cs}_i$ is the input for generating the queries. In addition, the keys and values are generated by the weighted-sum of contributions from the other channels, $\{\hat{H}^{cs}_j\}^{C}_{j=1, j \ne i}$, i.e. Eq.~(\ref{eq:h_cca}), which is similar to the beamforming process. Note that $A_j$, $W^{cc,*}$ and $\textbf{b}^{cc,*}$ are learnable weight and bias parameters, and $\odot$ indicates element-wise multiplication. The cross-channel attention output is then computed by:
\begin{align}
    H^{cc}_i = Softmax\left(\frac{Q^{cc}_i (K^{cc}_i)^T}{\sqrt{d_{m}}}\right) V^{cc}_i
    \label{eq:cca}
\end{align}
To the best of our knowledge, it is the first time this cross channel attention mechanism is introduced within the transformer network for multi-channel ASR.

Similar to CSA, we feed the contextual channel-attended representations through feed forward layers to get the final cross-channel attention outputs, $\hat{H^{cc}_i}$, as shown on the top of Fig.~\ref{fig:attentions}(c).
To learn more sophisticated contextual representations, we stack multiple CSAs and CCAs to from the encoder network output $\{H^e_i\}^C_{i=1}$ in Fig.~\ref{fig:attentions}(d).

\subsection{Multi-channel Decoder}
\textbf{Multi-channel encoder-decoder attention (EDA)}: Similar to~\cite{vaswani2017attention}, we employ the masked self-attention layer (MSA), $\hat{H^{sa}}$, to model the contextual relationship between target tokens and their predecessors. It is computed similarly as in Eq.~(\ref{eq:cs_qkv}) and (\ref{eq:csa}) but with token embeddings (Eq.~\ref{eq:token_embedding}) as inputs. Then we create the queries by $\hat{H^{sa}}$, and keys as well as values by the multi-channel encoder outputs $\{H^e_i\}^C_{i=1}$ as follows:
\begin{align}
    & Q^{ed} = \sigma\left(\hat{H^{sa}} W^{ed,q} + \textbf{1} (\textbf{b}^{md,q})^T\right) \notag \\
    & K^{ed} = \sigma\left(\frac{1}{C}\sum^{C}_{i=1} H^e_i W^{ed,k} + \textbf{1} (\textbf{b}^{md,k})^T\right) \label{eq:md_qkv} \\
    & V^{ed} = \sigma\left(\frac{1}{C}\sum^{C}_{i=1} H^e_i W^{ed,v} + \textbf{1} (\textbf{b}^{ed,v})^T\right) \notag
\end{align}
Again, $W^{ed,*}$ and $\textbf{b}^{ed,*}$ are learnable weight and bias parameters. The multi-channel decoder attention then becomes the regular encoder-decoder attention of the transformer decoder. Similarly, by applying MH-SDPA, layernorm, and feed forward layer, we can get final decoder output, $\hat{H^{ed}}$, as shown on the top of Fig.~\ref{fig:attentions}(d).
To train our multi-channel transformer, we use the cross-entropy loss with label smoothing of value $\epsilon_{ls}=0.1$~\cite{szegedy2016rethinking}.

\section{Experiments}
\label{sec:exps}

\begin{comment}
\begin{table*}[t]
\begin{center}
\caption{The relative word error rate reduction, WERRs (\%), by comparing all the multi-channel based methods to the single channel based transformer and SD beamformer cascaded with transformer (the numbers in parentheses). The higher the value is the better.}
\label{tab:mct}
\begin{tabular}{l|@{\hskip 1mm}c@{\hskip 1mm}|@{\hskip 1mm}c@{\hskip 1mm}|@{\hskip 1mm}c@{\hskip 1mm}} \hline
Method & No. of channels & No. of parameters & WERR (\%) over SCT \\
& &  (Million) & (over SDBF-T) \\ \hline
Single channel + Transformer (SCT) & 1 & 13.29 & - (-)  \\
Super-directive beamformer \cite{doclo2007superdirective} + Transformer (SDBF-T) & 7 & 13.29 & 6.27 (-)  \\ \hline
Neural beamformer \cite{kumatani2019multi} + Transformer (NBF-T) & 2 & 13.31 & 2.42 (-4.11) \\
Neural masked-based beamformer \cite{heymann2016neural} + Transformer (NMBF-T) & 2 & 18.53 & 2.07 (-4.49)  \\
Multi-channel Transformer with 2 channels (MCT-2) & 2 & 13.63 & \textbf{11.21} (\textbf{5.26}) \\ \hline
Multi-channel Transformer with 3 channels (MCT-3) & 3 & 13.80 & \textbf{20.70} (\textbf{15.39}) \\ \hline
\end{tabular}
\end{center}
\vskip -10pt
\end{table*}
\end{comment}
\begin{table*}[t]
\begin{center}
\caption{The relative word error rate reduction, WERRs (\%), by comparing the multi-channel transformer (MCT) to the beamfomers cascaded with transformers. A higher number indicates a better WER.}
\label{tab:mct}
\begin{tabular}{l|@{\hskip 1mm}c@{\hskip 1mm}|@{\hskip 1mm}c@{\hskip 1mm}|@{\hskip 1mm}c@{\hskip 1mm}|@{\hskip 1mm}c@{\hskip 1mm}|@{\hskip 1mm}c@{\hskip 1mm}|@{\hskip 1mm}c@{\hskip 1mm}} \hline
Method & No. of & No. of parameters & WERR over & WERR over & WERR over & WERR over\\
& channels &  (Million) & SCT & SDBF-T & NBF-T & NMBF-T \\ \hline
SC + Transformer (SCT) & 1 & 13.29 & - & - & - & - \\
SDBF \cite{doclo2007superdirective} + Transformer (SDBF-T)  & 7 & 13.29 & 6.27 & - & - & - \\ \hline
NBF \cite{kumatani2019multi} + Transformer (NBF-T) & 2 & 13.31 & 2.42 & -4.11 & - & -\\
NMBF \cite{heymann2016neural} + Transformer (NMBF-T) & 2 & 18.53 & 2.07 & -4.49 & - & - \\
MCT with 2 channels (MCT-2) & 2 & 13.63 & \textbf{11.21} & \textbf{5.26} & \textbf{9.00} & \textbf{9.33} \\ \hline
MCT with 3 channels (MCT-3) & 3 & 13.80 & \textbf{20.70} & \textbf{15.39} & \textbf{18.73} & \textbf{19.03}\\ \hline
\end{tabular}
\end{center}
\vskip -10pt
\end{table*}

\subsection{Dataset}
To evaluate our multi-channel transformer method (MCT), we conduct a series of ASR experiments using over 2,000 hours of speech utterances from our in-house anonymized far-field dataset. The amount of training set, validation set (for model hyper-parameter selection), and test set are 2,000 hours (312,0000 utterances), 4 hours (6,000 utterances), and 16 hours (2,5000 utterances) respectively. The device-directed speech data was captured using smart speaker with 7 microphones, and the aperture is 63mm. The users may move while speaking to the device so the interaction with the devices were completely unconstrained. In this dataset, 2 microphone signals of aperture distance and the super-directive beamformed signal by \cite{doclo2007superdirective} using 7 microphone signals are employed through all the experiments.

\subsection{Baselines}
We compare our multi-channel transformer (MCT) to four baselines: (1) \textbf{Single channel + Transformer (SCT)}: This serves as the single-channel baseline. We feed each of two raw channels individually into the transformer for training and testing, and obtain the average WER from the two channels. (2) \textbf{Super-directive (SD) beamformer \cite{doclo2007superdirective} + Transformer (SDBF-T)}:
The SD BF is widely used in the speech-directed devices including the one we used to obtain the beamformed signal in the in-house dataset. This beamformer used all seven microphones for beamforming. 
Multiple beamformers are built on the frequency domain toward different look directions and one with the maximum energy is selected for the ASR input; therefore, the input features to the transformer are extracted from a single channel of beamformed audio. (3) \textbf{Neural beamformer \cite{kumatani2019multi} + Transformer (NBF-T)}:
This serves as the fixed beamformer (FBF) baseline using two microphone signals as inputs rather than seven in SD beamformer. Multiple beamforming matrices toward seven beam directions followed by a convolutional layer are learned to combine multiple channels, and then the energy features from all beam directions respectively. The beamforming matrices are initialized with MVDR beamformer \cite{capon1969high}. 
(4) \textbf{Neural masked-based beamformer \cite{heymann2016neural} + Transformer (NMBF-T)}: It serves as the adaptive beamforming (ABF) baseline, and also uses two microphone signals as inputs. The mask estimator was pre-trained following \cite{heymann2016neural}. Note that the above neural beamforming models are jointly finetuned with the transformers.

\subsection{Experimental Setup and Evaluation Metric}
The transformers in all the baselines and our multi-channel transformer (MCT) are of $d_{m}=256$, number of hidden neurons $d_{ff}=1,024$, and number of heads, $h=3$. While MCT and the transformer for NMBF-T have $N_e=4$ and $N_d=4$, other transformers are of $N_e=6$, $N_d=6$ in order to have comparable model size, as shown in Table~\ref{tab:mct}. Note that NMBF-T is about 5M larger than the other methods due to the BLSTM and FeedForward layers used in the mask estimator of \cite{heymann2016neural}. Results of all the experiments are demonstrated as the relative word error rate reduction (WERR). Given a method A's WER ($\text{WER}_A$) and a baseline B's WER ($\text{WER}_B$), the WERR of A over B can be computed by $(\text{WER}_B - \text{WER}_A) / \text{WER}_B$; the higher the WERR is the better.

The input features, the Log-STFT square magnitude (for SCT and SDBF-T) and STFT (for NBF-T and NMBF-T) are extracted every 10 ms with a window size of 25 ms from 80K audio samples (results in $T=166$ frames per utterance); the features of each frame is then stacked with the ones of left two frames, followed by downsampling of factor 3 to achieve low frame rate, resulting in $F=768$ feature dimensions. In the proposed method, we use both log-STFT square magnitude features, and phase features following \cite{wang2018combining,wang2018multi} by applying
the sine and cosine functions upon the principal angles of the STFT at each time-frequency bin. We used
the Adam optimizer \cite{kingma2014adam} and varied the learning rate following \cite{vaswani2017attention,dong2018speech} for optimization. The subword tokenizer \cite{sennrich-etal-2016-neural} is used to create tokens from the transcriptions; we use $L=4,002$ tokens in total.

\begin{table}[t]
\begin{center}
\caption{The WERRs (\%) over MCT (with both CSA and CCA) while using CSA only or CCA only.}
\label{tab:cca}
\begin{tabular}{l|@{\hskip 1mm}c@{\hskip 1mm}|@{\hskip 1mm}c@{\hskip 1mm}} \hline
Channel-wise  & Cross-channel & WERR (\%) \\ 
self attention (CSA) & attention (CCA) & over MCT \\ \hline
\hspace{7mm}\cmark & \cmark & 0 \\
\hspace{7mm}\cmark & \xmark & -12.71 \\
\hspace{7mm}\xmark & \cmark & -13.12 \\ \hline
\end{tabular}
\end{center}
\vskip-10pt
\end{table}

\subsection{Experimental Results}
Table~\ref{tab:mct} shows the performances of our method (MCT-2) and beamformers+transformers methods over different baselines. While all cascaded beamformers+transformers methods perform better than SCT (by 2.07\% to 6\%), our method improves the WER the most (by 11.21\%). When comparing WERRs over SDBF-T, however, only MCT-2 improves the WER. The degradations from NBF-T and NMBF-T over SDBF-T may be attributed to not only 2 rather than 7 microphones are used but also the suboptimal front-end formalizations either by using a fixed set of weights for look direction fusion (NBF-T) or flawed speech/noise mask estimations (NMBF-T). 
If we compare our method directly to NBF-T and NMBF-T, we see 9\% and 9.33\% relative improvements respectively.
We further investigate whether the information from the super-directive beamformer channel was complementary to the multi-channel transformer. To this end, we take the beamformed signal from SD beamformer as the third channel and feed it together with the other two channels to our transformer (MCT-3). We see in Table~\ref{tab:mct} (the last row), about 10\% extra relative improvements are achieved compared to MCT-2.

\begin{figure}[t]
\centering
\includegraphics[width=0.43\textwidth]{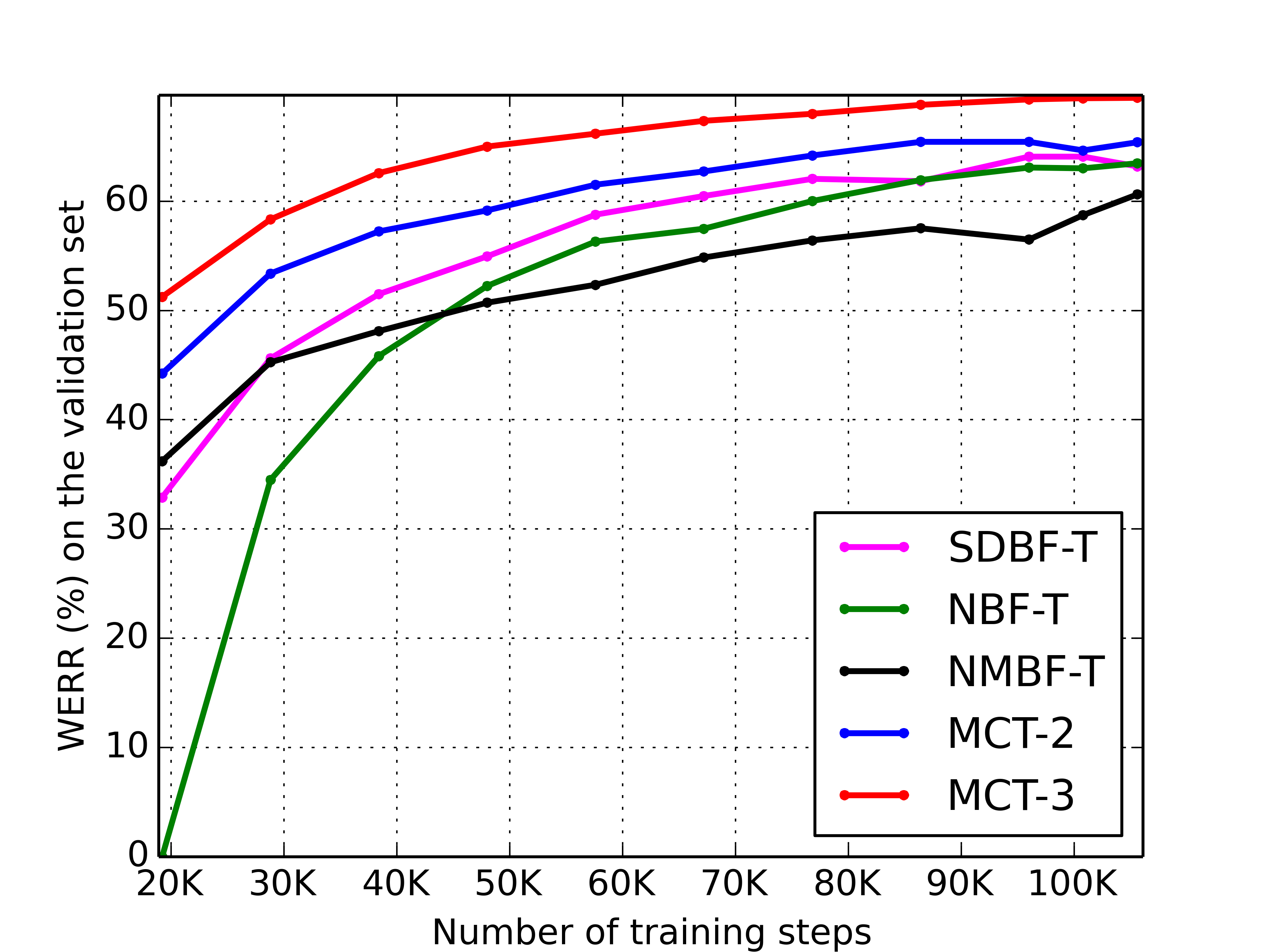}
\caption{The WERR w.r.t. the training steps of our methods (MCT-2,3) comparing to beamformers cascaded with transformers. Our model has started to converge at around 100K steps, but not for the others.}
\label{fig:devwers_mct}
\end{figure}

In Fig.~\ref{fig:devwers_mct}, we evaluate the convergence rate and quality via comparing the learning curves of our model to the other beamformer-transformer cascaded methods. Note that our model has started to converge at around 100K training steps, while the others have not. We compute the WERRs of all methods over a fixed reference point, which is the highest WER point during this period by NBF-T (the left-most point of NBF-T corresponding to WERR=0). Our method converges faster than the others with consistently higher relative WER improvements. Also, we observe NMBF-T converges the slowest, and the NBF-T is the second slowest. 

Finally, we conducted an ablation study to demonstrate the importance of channel-wise self attention (CSA) and cross-channel attention (CCA) layers. To this end, we train two variants of multi-channel transformers by using CSA only or CCA only. Table~\ref{tab:cca} shows that the WERR drops significantly when either attention is removed. 

Furthermore, our model can be simply applied on more than 3 channels. In an 8-microphone case, the number of parameters would increase by only about 10\% ($T \times d_m \times N_e \times 8/10^6/13.3=166\times256\times4\times8/10^6/13.3$) compared to the one-microphone case ($13.3$M parameters).

\begin{comment}
\begin{figure}[t]
\centering
\includegraphics[width=0.48\textwidth]{figs/Ablation.png}
\caption{The validation WERs (\%) of our model with channel-wise attention only, with cross-channel attention only, and with both w.r.t. the number of training steps.}
\vspace{-4mm}
\label{fig:ablation_mct}
\end{figure}
\end{comment}

\begin{comment}
\begin{figure}[t]
\centering
\includegraphics[width=0.43\textwidth]{figs/variants.png}
\caption{The WERRs (\%) of the proposed model with three different variations (please see texts for variant definitions).}
\vspace{-4mm}
\label{fig:variants_mct}
\end{figure}
\end{comment}

\begin{comment}
\begin{table}[t]
\begin{center}
\small
\begin{tabular}{l|@{\hskip 1mm}c@{\hskip 1mm}|@{\hskip 1mm}c@{\hskip 1mm}} \hline
Method & No. channels & WERR (\%)  \\ \hline
Single channel + Transformer & 1 & -  \\
SD BF \cite{doclo2007superdirective} + Transformer & 7 & 5.34 (-)  \\
Multi-channel transformer (Ours) & 2 & 9.86 (4.78) \\
Multi-channel transformer (Ours) & 3 & \textbf{19.91} (\textbf{15.39}) \\ \hline
\end{tabular}
\end{center}
\caption{The effect of different number of channels for the proposed multi-channel transformer against the single channel based transformer and SD beamformer cascaded with transformer (the numbers in the parentheses). The higher the value is the better.}
\label{tab:numchannels}
\end{table}
\end{comment}

\section{Conclusion}
\label{sec:conc}

We proposed an end-to-end transformer based multi-channel ASR model. We demonstrated that our model can capture the contextual relationships within and across channels via attention mechanisms. The experiments showed that our method (MCT-2) outperforms three cascaded beamformers plus acoustic modeling pipelines in terms of WERRs, and can be simply applied to more than 2 channel cases with affordable increases of model parameters.

{\small
\bibliographystyle{IEEEbib}
\bibliography{references}
}

\end{document}